\documentclass[prb,twocolumn,showpacs,amsmath,amssymb]{revtex4}

\usepackage[dvips]{graphicx}
\usepackage{bm}

\begin{document}

\title{Anomalous Hall conductivity: local orbitals approach}

\author{Pavel St\v{r}eda}

\affiliation{Institute of Physics, Academy of Sciences of the
Czech Republic, Cukrovarnick\'{a} 10, 162 53 Praha, Czech Republic}

\date{\today}

\begin{abstract}
A theory of the anomalous Hall conductivity based on
the properties of single site orbitals is presented.
Effect of the finite
electron life time is modeled by energy fluctuations of
atomic-like orbitals. Transition from the ideal Bloch system
for which the conductivity is determined by the Berry phase
curvatures to the case of strong disorder for
which the conductivity becomes dependent on the relaxation time
is analyzed. Presented tight-binding model gives by the
unified way experimentally observed qualitative features
of the anomalous conductivity in the so called good metal
regime and that called as bad metal or hopping regime. 
\end{abstract}


\maketitle

\section{Introduction}
It has been known for more than a century that ferromagnetic materials
exhibit, in addition to the standard Hall effect when placed in
a magnetic field, an extraordinary Hall effect which does not vanish
at zero magnetic field. The theory of this so-called anomalous Hall
effect has a long and confusing history, with different approaches
giving in some cases conflicting results.
While more recent calculations have somewhat unified the different
approaches and clarified the situation, it is still an active topic
of research, see for example the recent review by Nagaosa et al
\cite{Nagaosa2009}).

It is generally accepted that
anomalous Hall effect is induced by spin-orbit coupling.
It was first suggested by Karplus and Luttinger \cite{Karplus}
in 1954 to explain anomalous Hall effect observed on
ferromagnetic crystals.
Their analysis leads to the scattering independent
off-diagonal components of the conductivity,
which are attributed to the so-called ``intrinsic'' effect.
Later, theories of this effect based on several specific
models have been developed \cite{Miyazawa,Onoda}.
As shown recently, it is accompanied by
a strong orbital Hall effect \cite{Shindou,Kontani_1}.
The conductivity is affected by scattering,
which in the presence of spin-orbit coupling is basically of
two types, the so called side-jump \cite{Berger} and skew scattering
\cite{Smit,Luttinger_58,Sinitsyn}. They also
lead to an anomalous Hall effect, called ``extrinsic''.
It has also been argued that in the anomalous Hall
regime a periodic field of electric dipoles
(electric polarizability) is induced by the applied
current \cite{Karplus,Adams,Fivaz}. 
This property has recently been shown to be related to the so-called
orbital polarization moment \cite{Streda_Jonck} which
is determined by Berry phase curvature in pure systems \cite{Niu}.

The best quantitative agreement with experimental observations
has been obtained by semi-classical transport theory
\cite{Niu}, leading to the Berry phase correction to the group
velocity. For Fe crystals \cite{Jungw_2} it yields an anomalous
conductivity $\sim 750 \; \Omega^{-1}$ cm$^{-1}$  while
a value approaching $ 1000 \; \Omega^{-1}$ cm$^{-1}$
has been observed.
However, up to now, generalization of this theory to
systems with strong disorder or subject to other types
of fluctuations seemed to fail. It is the main aim
of this paper to present theoretical treatment filling
this gap.

In contrast to standard transport theories the Hall conductivity
is expressed in terms of local atomic-like orbitals.
It is explicitly derived that for perfect Bloch electron systems
this description coincides with that given by Berry phase curvatures
indicating intrinsic, topological, origin of anomalous Hall
effect. Presented view based on atomic-like orbitals
allows to include effect of disorder by a local energy fluctuation
of these orbitals which is an alternative description of
scattering events. Fluctuations modify the Hall conductivity
and transition from perfect to strongly disordered systems
is analyzed. So called "intrinsic" and "extrinsic"
Hall effect are just treated by an unified way. To test
presented view based on local atomic-like orbitals  
the two band model within tight-binding approach is used.
Obtained dependence of the anomalous Hall conductivity
on the relaxation time shows the observed qualitative
features \cite{Miyasato}. Similar features has also been
obtained by the different procedure with relaxation
time being a fitting parameter \cite{Kontani_2}.
In contrast to this work the presented model allows
to relate relaxation time to fluctuations of atomic-like
orbitals, i.e. to its microscopic origin.
By the unified way it gives experimentally
observed scaling of anomalous Hall conductivity $\sigma_{xy}$
with diagonal conductivity component $\sigma_{xx}$ in
the region of the so called good metal for which
$\sigma_{xy} \sim (\sigma_{xx})^0$ and that of the bad
metal for which $\sigma_{xy} \sim (\sigma_{xx})^{1.6}$.

The paper is organized as follows.
In section II basic properties of electron systems in crystalline
structures with spin-orbit coupling are summarized, while
basic expressions for Hall conductivity derived by using
the quantum linear response theory are rederived in
the section III. In section IV an alternative expression
for Hall conductivity, including on-energy shell matrix elements
only, is derived. Section V is devoted to perfect Bloch electron systems
where conductivity is expressed in terms of local orbital
polarization moments which are further expressed via the Berry phase
curvature. In section VI a two-band model in tight-binding
approach is presented to estimate the effect
of a finite electron life time to the anomalous
Hall conductivity at zero magnetic field.
Presented theory of anomalous Hall effect is
summarized and commented in the last section.

\section{Single electron Hamiltonian and statistical fluctuations}
Within a mean field approach, electron properties are controlled
by a single electron Hamiltonian which we consider
in the following standard form
\begin{eqnarray} \label{Hamiltonian}
H \, = \, \frac{\vec{p}^{\; 2}}{2m_0} \, + \,
V(\vec{r}) \, + \,
\frac{\lambda_c^2}{4 \hbar} \; \vec{\sigma} \cdot
\left[ \vec{\nabla}V(\vec{r}) \, \times \, \vec{p} \, \right] \, + \,
\nonumber \\ - \,
\mu_B \, \vec{B}_{\rm eff} \cdot \vec{\sigma}
\; .
\end{eqnarray}
Here, $m_0$ and $\mu_B$ denote free electron mass and
the Bohr magneton, respectively, $\vec{p}$ is the momentum operator,
$V(\vec{r})$ denotes a background potential
and components of $\vec{\sigma}$ are Pauli matrices.
The third term on the right hand side represents
spin-orbit coupling with $\lambda_c$ being an effective
Compton length. The last term on the right hand side
describes Zeeman-like spin splitting due to the exchange-correlation
energy represented by an effective field $\vec{B}_{\rm eff}$
which can generally be position-dependent.
The corresponding velocity operator reads
\begin{equation} \label{v_def_so}
\vec{v} \, = \, \frac{1}{i \hbar} \,
[ \, \vec{r} \, , \, H \, ] \, = \,
\frac{\vec{p}}{m_0} \, + \, \frac{\lambda_c^2}{4 \hbar} \;
\vec{\sigma} \times \vec{\nabla} V(\vec{r})
\; .
\end{equation}

Strictly speaking, Hamiltonian $H$ quite well defines
properties of electrons located within a finite volume of
characteristic dimensions determined only by the 
electron coherence
length. Within each of such volumes the background potential
as well $\vec{B}_{\rm eff}$ can be different. This way,
a set of electron systems, the statistical ensemble, is defined.
If time-dependent fluctuations can be treated
within an adiabatic approach \cite{R1_adiabatic_approach},
they can also be included in this ensemble.
Measurable quantities are given by their
statistically averaged values. It is useful to split
Hamiltonian $H$, Eq.~(\ref{Hamiltonian}), into two parts
\begin{equation}
H \, = \, H_0 \, + \, \delta H
\qquad , \qquad
H_0 \, \equiv \, \langle H \rangle_{\rm av}
\; , 
\end{equation}
where only $\delta H$ depends on statistical fluctuations.
For crystalline solids, to which the present treatment is devoted,
statistically averaged Hamiltonian $H_0$,
obeys full crystal symmetry and the effective field $\vec{B}_{\rm eff}$
can be assumed constant. It defines the so 
called virtual crystal with eigenstates $|n, \vec{k} \rangle$
of the energy $E_n(\vec{k})$, characterized by the band index $n$
and the wave vector $\vec{k}$.
Eigenfunctions are two-component Bloch spinors
\begin{eqnarray} \label{eigenstates}
|n , \vec{k} \, \rangle \, & = & \, \Psi_{n,\vec{k}}(\vec{r}) \, = \,
\frac{e^{i \vec{k} \vec{r}}}{\sqrt{8 \pi^3}} \,
u_{n , \vec{k}}(\vec{r})
\; , \\
H_0 \, |n , \vec{k} \, \rangle \, & \equiv & \,
\langle H \rangle_{\rm av} \, |n , \vec{k} \, \rangle \, = \,
E_n(\vec{k}) \, |n , \vec{k} \, \rangle
\; , \nonumber
\end{eqnarray}
where $u_{n, \vec{k}}(\vec{r})$ are spinor functions periodic in
$\vec{r}$ with period defined by the elementary lattice
translations. Velocity matrix elements are diagonal in the
wave vector $\vec{k}$ located within the Brillouin zone
\begin{equation} \label{v_n,n'_k}
\langle n , \vec{k} | \vec{v} \, | n' , \vec{k}'\, \rangle \, = \,
\vec{v}^{\; n, n'}(\vec{k}) \; \delta_{\vec{k}, \vec{k}'}
\; ,
\end{equation}
and the expectation values read
\begin{equation}
\vec{v}_{n}(\vec{k}) \, \equiv \,
\langle n , \vec{k} | \vec{v} \, | n , \vec{k}\, \rangle \, = \,
\frac{1}{\hbar} \, \vec{\nabla}_{\vec{k}} E_n(\vec{k})
\; .
\end{equation}

Equilibrium properties are determined
by the effective Hamiltonian, $H_{\rm eff}$, defined by
the statistically averaged Green's function
\begin{equation} \label{H_eff}
\langle G(z) \rangle_{\rm av} \, \equiv \,
\left\langle \frac{1}{z - H} \right\rangle_{\rm av} \, \equiv \,
\frac{1}{z - H_{\rm eff}(z)}
\; ,
\end{equation}
where $z$ is the complex energy variable. It has the full
crystal symmetry and it is diagonal in the representation
given by eigenstates of the averaged Hamiltonian $H_0$,
Eq.~(\ref{eigenstates}). 
Effective Hamiltonian is non-Hermitian and energy dependent
but it is analytic in both complex half-planes,
$H_{\rm eff}(z^{\ast}) \, = \, H_{\rm eff}^{+}(z)$.
Its standard form reads
\begin{equation} \label{self-energy}
H_{\rm eff}(z) \, = \, H_0 \, + \,
\Sigma(z) \quad , \quad
\Sigma(z) \, = \, \Delta(z) - i \Gamma(z)
\; ,
\end{equation}
where  $\Sigma(z)$ is the energy dependent self-energy. Inverse
value of its imaginary part represents a finite electron
life-time. 

To include an external magnetic field $\vec{B}$ the Hamiltonian
defined by Eq.~(\ref{Hamiltonian}) has to be modified. Both
the momentum operator entering the Hamiltonian, and the velocity
operator, Eq.~(\ref{v_def_so}), have to be
replaced by their counterparts, which include a vector potential $\vec{A}$.
\begin{equation}
\vec{p} \, \rightarrow \,
\vec{p} \, + \, e \vec{A}
\; .
\end{equation}
Here, $e$ denotes the electron charge absolute value and the magnetic
field is given as  $\vec{B} = {\rm curl} \vec{A}$.
Also the value of the parameter $\vec{B}_{\rm eff}$ defining
Zeeman-like splitting is modified by $\vec{B}$.
The external magnetic field generally removes translation symmetry.
Exceptions are the so called rational magnetic fields for which
the problem becomes invariant under translations with
different elementary translations than those 
given by the periodic potential.

\section{Linear response theory}
In this section the standard linear response theory will be
described and well known general formulae derived
to summarize basic theoretical assumptions.
In accord with the original work by Kubo \cite{Kubo}
the external electric field is supposed to be turned on
at the time $t \rightarrow -\infty$ and reach the final value
$\vec{\cal{E}}$ at $t=0$. Exponential time evolution is
considered, 
$\vec{\cal{E}}(t) = \vec{\cal{E}} \, e^{\epsilon t/\hbar}$,
with $\epsilon>0$ being an infinitesimally small
quantity. It gives rise to the Hamiltonian perturbation
$e \vec{r} \cdot \vec{\cal{E}}(t)$ and corresponding
density matrix ${\bm \rho}(t)$ has to satisfy the equation
of motion which reads
\begin{equation}
\frac{\partial {\bm \rho}(t)}{\partial t} \, + \, \frac{1}{i \hbar}
\left[{\bm \rho}(t) \, , \, H +
e \vec{r} \cdot \vec{\cal{E}}(t)
\right] \, = \, 0
\; .
\end{equation}
It can be written in the following way
\begin{equation}
{\bm \rho}(t)  \, = \, {\bm \rho}_0(H) \, + \,
e^{-i H t/ \hbar} {\bm \rho}_1(t) \, e^{i H t/ \hbar}
\; ,
\end{equation}
where ${\bm \rho}_1(t)$ represents the deviation
from the density matrix in the absence of electric field
given by the Fermi-Dirac distribution function
\begin{equation}
{\bm \rho}_0(H) \, = \,
\left( 1 + e^{\frac{H-\mu}{k_{\rm B} T}} \right)^{-1}
\; ,
\end{equation}
where $\mu$ and $k_{\rm B} T$ are the chemical potential
and thermal energy, respectively.
Up to linear terms in the electric field, ${\bm \rho}_1(t)$
reaches the following value at $t=0$
\begin{equation} \label{rho_1}
{\bm \rho}_1(0) =
\frac{i}{\hbar}
\int\limits_0^{+ \infty} e^{- \epsilon t / \hbar}
e^{- i H t / \hbar}
\left[ {\bm \rho}_0(H) \, , \, e \vec{r} \cdot \vec{\cal{E}} \right]
e^{i H t / \hbar} dt
\; ,
\end{equation}
and the resulting current density reads
\begin{equation} \label{vec_j_lim}
\vec{j} \, = \, - \, e \, \lim_{\epsilon \rightarrow 0^+}
\left\langle \,
{\rm Tr} \left[\, {\bm \rho}_1(0) \, \vec{v} \, \right] \,
\right\rangle_{\rm av}
\; .
\end{equation}

Generally, the above potential perturbation
$e \vec{\cal{E}} \vec{r}$ of the Hamiltonian is not the only
perturbation caused by electric field. The potential
gradient $e \vec{\cal{E}}$ enters the spin-orbit term
of the Hamiltonian, Eq.~(\ref{Hamiltonian}), as well as
the velocity operator, Eq.~(\ref{v_def_so}), giving rise
to additive terms linear in electric field. Resulting
contributions to the current density are of a
higher order than $\lambda_c^2$ in the Compton length.
For the considered Hamiltonian $H$,
Eq.~(\ref{Hamiltonian}), which includes spin-orbit
coupling only approximately up to the order of $\lambda_c^2$,
these contributions thus have to be ignored.

Introducing the $\delta$-function operator
\begin{eqnarray} \label{G_pm}
\delta(\eta -H) = - \lim_{\epsilon \rightarrow 0^+} \,
\frac{G^+(\eta) -G^-(\eta)}{2 \pi i}
\; \; ,
\\
G^{\pm}(\eta) = \frac{1}{\eta - H \pm i \epsilon}
\, , \nonumber
\end{eqnarray}
the time integration, Eq.~(\ref{rho_1}), can easily be performed.
For the limiting case of the zero temperature in 
function $\rho_0(H)$ the components of the conductivity
tensor obey the following form
\begin{eqnarray} \label{sigma_ij_1}
\sigma_{i j}(\mu) \equiv
\frac{j_i}{{\cal{E}}_j} \, = \, \qquad \qquad \qquad
\nonumber \\ =
- e^2 \! \! \! \int\limits_{-\infty}^{\mu} \! \! \!
\left\langle {\rm Tr} \left\{ \delta(\eta - H) \left[
v_{i} G^+ r_{j} +
r_{j} G^- v_{i} \right]
\right\} \right\rangle_{\rm av} \! d \eta
\, ,
\end{eqnarray}
where $i,j = x,y,z$.
The proper way to treat the limits in Eqs.~(\ref{vec_j_lim})
(due to the electric field time evolution) and (\ref{G_pm}) 
(regularization of $\delta(H-\eta)$) would be to introduce 
two different infinitesimal parameters and treat both of them
independently, after all other steps are taken. However, 
our procedure of statistical averaging yields a
non-zero imaginary part of the self-energy, $\Gamma$,
entering both averaged operators, and the limiting case
of the fully coherent system is defined by the physically
acceptable limit $\Gamma \rightarrow 0^+$.
That is why it is sufficient to consider just one infinitesimally
small parameter $\epsilon \rightarrow 0^+$
implicitly entering Eq.~(\ref{sigma_ij_1}).
Misunderstanding of this limit has been reason for
doubts concerning validity of resulting expressions, especially
of the Hall conductivity.

The non-zero temperature smearing effect of the distribution
function yields
\begin{eqnarray} \label{sigma_ij_T}
\sigma_{i j}(\mu ,T) \, = \, -
\int\limits_{-\infty}^{+\infty}
\frac{d {\bm \rho}_0(\eta)}{d \eta} \;
\sigma_{i j}(\eta) \, d \eta
\; .
\end{eqnarray}
This relation allows to limit our attention to the
analysis of the energy-dependent conductivity given
by Eq.~(\ref{sigma_ij_1}). 
It does not mean that $\sigma_{ij}(\mu)$ has to be
temperature independent since potential fluctuations
as well as some of the Hamiltonian parameters
can be implicitly temperature-dependent quantities,
as is e.\ g.\ $\vec{B}_{\rm eff}$ representing 
the exchange-interaction effect.

Using definition of the velocity operator
\begin{equation} \label{v_def_general}
\vec{v} \, = \, \frac{1}{i \hbar} \, \left[ \vec{r} , H \right] = -
\frac{1}{i \hbar} \,
\left[ \vec{r} , \left( G^{\pm}(\eta) \right)^{-1} \right]
\; ,
\end{equation}
and the identity
\begin{equation}
\left[ G^{\pm}(\eta) \right]^2 \, = \, - \,
\frac{d G^{\pm}}{d \eta}
\; ,
\end{equation}
the Eq.~(\ref{sigma_ij_1}) can be rewritten \cite{remark_2}
in a form including velocity operators only
\begin{eqnarray} \label{Bastin}
\sigma_{i j}(\mu) = i \hbar e^2 \times
\qquad \qquad \qquad \qquad \qquad \qquad \qquad \quad
\nonumber \\ \times \! \! \!
 \int\limits_{-\infty}^{\mu} \! \!
\left\langle {\rm Tr}
\left\{ \delta(\eta - H) \left[
v_{i} \frac{d G^{+}}{d \eta} v_{j} -
v_{j} \frac{d G^{-}}{d \eta} v_{i} \right]
\right\} \right\rangle_{\rm av} \! \! d \eta
\; . \;
\end{eqnarray}
It coincides with that obtained by Bastin et al
\cite{Bastin} by the use of a different procedure.
For diagonal components of the conductivity tensor the
integration by parts gives the well known Kubo-Greenwood
formula \cite{Greenwood}
\begin{equation} \label{sigma_ii}
\sigma_{i i}(\mu) \, = \, \pi e^2 \hbar
\left\langle \, {\rm Tr} \left\{
v_{i} \delta(\mu - H) v_{i} \delta(\mu - H)
\right\} \, \right\rangle_{\rm av}
\; .
\end{equation}
Making use of the following identity \cite{remark_2,Smrcka}
\begin{eqnarray}
i \hbar {\rm Tr}
\left\{ \delta(\eta - H) \left[
v_{i} \frac{d G^{+}}{d \eta} v_{j} -
v_{j} \frac{d G^{-}}{d \eta} v_{i} \right]
\right\} \, = \qquad
\nonumber \\ = \,
\frac{i \hbar}{2} \,
\frac{d}{d \eta} \,  {\rm Tr} \left\{ \delta(\eta - H)
\left[ v_{i} G^+(\eta) v_{j} - v_{j} G^-(\eta) v_{i}
\right] \right\} \, +
\nonumber \\ +  \frac{1}{2} \, \frac{d}{d \eta} \,
{\rm Tr} \left\{ \delta(\eta - H)
\left[ r_{i} v_{j} - r_{j} v_{i}
\right] \right\}
\; , \; \;
\end{eqnarray}
the off-diagonal conductivity components can be split into two
parts \cite{Streda_formula}
\begin{equation} \label{sigma_I_II}
\sigma_{xy}(\mu) =
\sigma_{xy}^{\rm (I)}(\mu) + \sigma_{xy}^{\rm (II)}(\mu) = -
\sigma_{yx}(\mu)
\; ,
\end{equation}
where
\begin{eqnarray} \label{sigma_I}
\sigma_{xy}^{\rm (I)}(\mu) = e^2 \,
\frac{i \hbar}{2} \, \times
\hspace{47mm}
\nonumber \\ \times
\left\langle {\rm Tr} \left\{
\delta(\mu - H) \left[ v_{x} G^+(\mu) v_{y} -
v_{y} G^-(\mu) v_{x} \right]
\right\} \right\rangle_{\rm av}
\; ,
\end{eqnarray}
and
\begin{eqnarray} \label{sigma_II}
\sigma_{xy}^{\rm (II)}(\mu) =  \frac{e^2}{2}
\left\langle {\rm Tr} \left\{ \delta(\eta - H)
\left[ x v_y - y v_x \right]
\right\} \right\rangle_{\rm av} \, =
\nonumber \\ = \, -
e^2 \left\langle {\rm Tr} \left\{ \delta(\eta - H) \,
y \, v_x \right\} \right\rangle_{\rm av}
\; .
\end{eqnarray}
These conductivity formulae are quite general, they also include both
the effect of the magnetic field and the spin-orbit coupling
\cite{Bruno}.
Similar treatment, but formally more complicated \cite{Smrcka},
is applicable to other transport coefficients,
like the thermopower and the heat conductivity.

The above expressions for the Hall conductivity $\sigma_{xy}(\mu)$
have been particularly useful for understanding of the quantum Hall effect.
While the contribution $\sigma_{xy}^{\rm (I)}(\mu)$ vanishes
whenever $\mu$ is located within an energy gap, the contribution
$\sigma_{xy}^{\rm (II)}(\mu)$ can remain finite, giving rise to
quantized values of the Hall conductivity \cite{Streda_formula}.
However, these formulae are not very convenient for $\mu$
located within an energy band since, at least for some specific
models \cite{Streda_84,Kontani_2}, a non-negligible part
of $\sigma_{xy}^{\rm (I)}(\mu)$ is nearly canceled by part
of the $\sigma_{xy}^{\rm (II)}(\mu)$. Thus it should be
useful to derive an alternative form of the expression
for the Hall conductivity, which should also allow a clear
physical interpretation.

\section{Hall conductivity in terms of on-energy-shell
matrix elements}
Conductivity can directly be measured on samples
with so called Corbino disc geometry. In the limiting case of
the large disc radii such samples can be approximated
by strips with a rectangular cross-section $L_y L_z$.
Using a proper choice of the vector potential, periodic
boundary conditions along $\hat{x}$ direction can be
considered on the length $L_x$. This geometry
allows to apply an electric field along $\hat{y}$ direction, and
establish $\sigma_{yy}$, and also the Hall current
along $\hat{y}$ direction and consequently $\sigma_{xy}$.
Under these boundary conditions the eigenvalue problem for one
particular Hamiltonian of the form given by
Eq.~(\ref{Hamiltonian}) can be solved, at least in principle. 
Obtained eigenstates $|\alpha \rangle$ of energy
$E_{\alpha}$ represent one of the systems belonging to
the considered statistical ensemble.
A corresponding contribution to the Hall conductivity
can be analyzed by using this $\alpha$-representation.

The term given by Eq.~(\ref{sigma_I}) reads
\begin{eqnarray}
\sigma_{xy}^{(I)}(\alpha,E) = e^2 \,
\frac{i \hbar}{2} \sum_{\alpha,\alpha'}
\delta(E - E_{\alpha}) \, \times \qquad \qquad
\nonumber \\ \times \left[
v_{x}^{\alpha, \alpha'} {G}^+_{\alpha'}(E)
v_{y}^{\alpha', \alpha} -
v_{y}^{\alpha, \alpha'} {G}^-_{\alpha'}(E)
v_{x}^{\alpha', \alpha}
\right]
\; ,
\end{eqnarray}
where $v_{x,y}^{\alpha, \alpha'}$ denotes velocity matrix
elements. Inserting commutation relation
Eq.~(\ref{v_def_general}) for the operator $v_y$, we get
\begin{eqnarray}
\sigma_{xy}^{(I)}(\alpha,E) = \frac{e^2}{2} \,
\sum_{\alpha,\alpha'}
\delta(E - E_{\alpha}) \,
(E_{\alpha} - E_{\alpha'} ) \, \times
\nonumber \\ \times \left[
\frac{v_{x}^{\alpha, \alpha'} y^{\alpha', \alpha}}
{E - E_{\alpha'} + i \epsilon} \, + \,
\frac{ y^{\alpha, \alpha'} v_{x}^{\alpha', \alpha}}
{E - E_{\alpha'} - i \epsilon} \right]
\; .
\end{eqnarray}
Note that because of the considered strip geometry, matrix
elements of the $y$ coordinate are finite.
Since terms for which $E_{\alpha} = E_{\alpha'}$
vanish, we get
\begin{eqnarray}
\sigma_{xy}^{(I)}(\alpha , E) =
\frac{e^2}{2} \, \sum_{\alpha,\alpha'}
\delta(E - E_{\alpha}) \times \qquad \qquad
\nonumber \\ \times
\left[ v_{x}^{\alpha, \alpha'} y^{\alpha', \alpha} +
y^{\alpha, \alpha'} v_{x}^{\alpha', \alpha} \right] \,
\left( 1 - \delta_{E_{\alpha'},E_{\alpha}} \right)
\; .
\end{eqnarray}
In $\alpha$-representation the remaining conductivity contribution
given by Eq.~(\ref{sigma_II}) can be written as
\begin{eqnarray}
\sigma_{xy}^{(II)}(\alpha, E) = - \frac{e^2}{2} \,
\sum_{\alpha,\alpha'}
\delta(E - E_{\alpha}) \, \times
\nonumber \\ \times
\left[ v_{x}^{\alpha, \alpha'} y^{\alpha', \alpha} +
y^{\alpha, \alpha'} v_{x}^{\alpha', \alpha} \right]
\; .
\end{eqnarray}
Sum of $\sigma_{xy}^{I}$ and $\sigma_{xy}^{II}$ yields the Hall 
conductivity expressed in terms of
on-energy-shell matrix elements only
\begin{equation} \label{sigma_xy_alpha}
\sigma_{xy}(E) = - e^2
\left\langle \sum_{\alpha,\alpha'}
\delta(E - E_{\alpha}) \,
v_{x}^{\alpha, \alpha'} y^{\alpha', \alpha} \,
\delta_{E_{\alpha'},E_{\alpha}}  \right\rangle_{\! \! \rm av}
\! \! , \; \; \; \;
\end{equation}
where averaging is taken over the $\alpha$-representations
of all elements of the considered ensemble.
In the case that the electron system
is fully coherent within the considered sample volume
$L_xL_yL_z$ the averaging procedure should be avoided.

The above expression for the Hall conductivity,
Eq.~(\ref{sigma_xy_alpha}), has clear interpretation
for quantum Hall effect. Let us assume that the Fermi energy
is located within region of localized states. From their
definition only diagonal matrix elements of the $y$
coordinate can differ from zero,
$y^{\alpha', \alpha} = y^{\alpha, \alpha} \delta_{\alpha', \alpha}$.
Since their velocity expectation values vanish,
contribution of localized states to the Hall conductivity
vanishes as has to be. 
Non-zero contribution can only be given by chiral
edge states giving rise to quantum Hall effect
non-affected by the presence of localized states.

\section{Perfect Bloch electron systems}
It is of particular interest to apply expression for the
Hall conductivity, Eq.~(\ref{sigma_xy_alpha}), to systems having
translation symmetry, including those modified by rational magnetic
fields. The basic property of these systems is that the velocity
matrix elements are diagonal in wave vector $\vec{k}$,
Eq.~(\ref{v_n,n'_k}). In general, the probability to find
on the Fermi surface states having the same $\vec{k}$
but belonging to different bands is statistically negligible,
except of the trivial case of band degeneracy.
Since Eq.~(\ref{sigma_xy_alpha}) includes only on-energy-shell
matrix elements, inter-band matrix elements do not
affect conductivity of ideal Bloch electron systems. 

To proceed further, let us first discuss the contribution 
to the Hall conductivity, Eq.~(\ref{sigma_xy_alpha}),
of states having zero velocity
expectation value along $\hat{y}$ direction.
It is given by 
\begin{equation}
e \, V_{\rm ws} \, \left( v_{x}^{n,n}(\vec{k}\,) \, y_n(\vec{k}\,)
\, + \,
v_{x}^{n,n}(-\vec{k} \, ) \, y_n(-\vec{k} \,) \right)
\; ,
\end{equation}
where $V_{\rm ws}$ denotes volume of the Wiegner-Seitz cell.
It is a contribution of the local orbital momentum to the conductivity.
As has been already discussed \cite{Streda_Jonck,polarizability_08}
it is responsible for charge polarization in transport
regime and that is why it can be called 
orbital polarization momentum.
States with non-zero velocity along $\hat{y}$ direction 
contribute by the product $-x v_y$ and the Hall conductivity
can be written as
\begin{equation}  \label{sigma_P_n}
\sigma_{xy}(E) \, = \, - \, \frac{e}{V_{\rm ws}} \,
\sum_n \left[ \vec{P}_n(E) \right]_z
\; ,
\end{equation}
where the orbital polarization momentum $\vec{P}_n(E)$ reads
\begin{equation} \label{OPM}
\frac{\vec{P}_{n}(E)}{V_{\rm ws}} \equiv -
\frac{e}{8 \pi^3}
\int\limits_{\rm BZ} \delta \left( E_n(\vec{k}) -E \right) \,
\vec{r}_n(\vec{k} \,) \times \vec{v}_n(\vec{k} \,)
\, d^3k
\; .
\end{equation}
Here $\vec{r}_n(\vec{k})$ denotes the expectation value
of the radius-vector of the orbital $n$ within the Wiegner-Seitz
cell and integration is limited 
to the Brillouin zone volume.

It can be easily shown that $\vec{P}_{n}(E)$ can be expressed in
terms of the Berry phase curvature. Since
\begin{equation}
\delta (E - E_n(\vec{k}) \,) \, = \, - \,
\frac{d f_0 (E_n(\vec{k}) \, )}{d \, E_n(\vec{k})}
\; ,
\end{equation}
where $f_0(E)$ denotes the zero-temperature Fermi-Dirac distribution,
we get
\begin{eqnarray}
\vec{P}_n(E) = -
\frac{e}{h} \frac{V_{\rm ws}}{4 \pi^2} \int\limits_{\rm BZ}
\left[ \vec{\nabla}_{\vec{k}} \, f_0(E_n(\vec{k}) \,) \right]
\times \vec{r}_n(\vec{k}) \,
d^3k
\, , \;
\end{eqnarray}
Integration by parts gives
\begin{eqnarray}
\vec{P}_n(E) = \frac{e}{h}
\frac{V_{\rm ws}}{4 \pi^2} \int\limits_{\rm BZ}
f_0 ( E_n(\vec{k}) \,) \,
\left( \vec{\nabla}_{\vec{k}}
\times \vec{r}_n(\vec{k}) \right) \,
d^3k
\, . \;
\end{eqnarray}
Using the expression derived in the Appendix for expectation values of 
the radius-vector
\begin{equation}
\vec{r}_n(\vec{k}) \, = \, - \,
\int\limits_{V_{\rm ws}} {\rm Im}
\left[ u^{+}_{n, \vec{k}}(\vec{r} \,) \, \vec{\nabla}_{\vec{k}} \,
u_{n, \vec{k}}(\vec{r} \,) \right] \, d^3r
\; ,
\end{equation}
we finally get
\begin{equation} \label{Bp}
\frac{\vec{P}_n(E)}{V_{\rm ws}} \, = \, \frac{e}{h}
\frac{1}{4 \pi^2} \int\limits_{\rm BZ}
f_0 ( E_n(\vec{k}) \,) \; \vec{\Omega}_n(\vec{k}) \, d^3k
\; ,
\end{equation}
where $\vec{\Omega}_n(\vec{k})$ is just the Berry phase
curvature \cite{Niu}
\begin{equation}
\vec{\Omega}_n(\vec{k}) = - {\rm Im}
\int\limits_{\rm V_{ws}} \! \!
\left( \vec{\nabla}_{\vec{k}} u^+_{n, \vec{k}}(\vec{r}) \right)
\times
\left( \vec{\nabla}_{\vec{k}} u_{n, \vec{k}}(\vec{r}) \right)
d^3r
\; .
\end{equation}
To conclude, the effect Berry phase curvatures is an
alternative description to the presented effect of the 
orbital polarization moment, Eq.~(\ref{OPM}).

\section{Tight-binding approach and single site fluctuations}
The aim of this section is to present a simple model system
allowing us to understand the main features of the anomalous Hall
conductivity at zero external magnetic field. For the sake
of simplicity the consideration will be limited to
isotropic systems, like those of the cubic symmetry.
A single band model Hamiltonian will be considered of the
following form
\begin{equation} \label{tb_Hamiltonian}
H \, = \, \sum_l | l \rangle
\left( E_a + \delta_l \right) \langle l |
\, + \,
\sum_{l,m}^{l \ne m} | l \rangle \, t_{l m} \, \langle m |
\; ,
\end{equation}
where $|l \rangle$ and $|m \rangle$ are Wannier functions
representing atomic-like orbitals associated with lattice
sites $\vec{R}_l$ and $\vec{R}_m$, respectively. To model
fluctuations, a variation $\delta_l$
of atomic-like orbital energies, $E_a$, will be considered,
while hopping integrals $t_{l m}$ are supposed
to be fluctuation independent quantities.
Instead of considering a specific form
of the band dispersion for averaged Hamiltonian,
$H_0 = \langle H \rangle_{\rm av} = H(\delta \rightarrow 0)$,
we shall assume it gives an elliptical density of states.
In accord with Hubbard \cite{Hubbard}, its energy
dependence normalized per Wiegner-Seitz volume $V_{\rm ws}$
can be written as follows
\begin{eqnarray}
\begin{array}{lc}
g(E) = 0 \; , &  |E-E_a| > w \; , \\
g(E) = \frac{2}{\pi w^2} \, \sqrt{ w^2 - (E-E_a)^2}
\; , & |E-E_a| \le w \; ,
\end{array}
\end{eqnarray}
where $2w$ denotes the band width. The corresponding mean
Fermi velocity reads
\begin{equation}
v_{\rm F} \, = \, \frac{w^2}{2 \hbar} \,
\left( \frac{\pi}{2} \right)^{2/3} \, \tilde{a} \, g(\mu)
\; ,
\end{equation}
where $\tilde{a} \equiv V_{\rm ws}^{1/3}$ just equals
to the lattice constant for the simple cubic lattice.
Note that this model system was already successfully applied
in a description of electron properties of substitutional alloys
\cite{Bedrich}. 
Under these simplifying assumptions the single site
orbital polarization moment defined by Eq.~(\ref{OPM})
becomes an energy dependent quantity, given as
\begin{equation} \label{P_l}
\vec{P}_l(E) = \vec{n}_a \tilde{a}^2
\frac{e}{h}\left( \frac{\pi}{2} \right)^{\frac{2}{3}}
\frac{r_{a}(\delta_l)}{\tilde{a}}
\left[ 1 - \left( \frac{E-E_{a} - \delta_l}{w} \right)^2 \right]
\; ,
\end{equation}
where $r_{a}(\delta_l)$ denotes the average radius of the considered
atomic-like orbital of the energy $E_a+\delta_l$ and
$\vec{n}_a$ is the unit vector parallel to its orbital
momentum. The only free model parameter in
$\vec{P}_l(E)$, and consequently in anomalous Hall conductivity,
Eq.~(\ref{sigma_P_n}), is the orbital radius. For
$\tilde{a} \sim 3$\AA $\,$ and $r_{a}(0)/\tilde{a} \sim 0.3$, the
corresponding anomalous Hall conductivity can reach values
of several hundred of $\Omega^{-1} {\rm cm}^{-1}$,
similar to those observed experimentally.

In real structures several overlapping energy bands contribute
to the conductivity. Let us for simplicity consider
two bands of the same width originated in atomic-like
orbitals of energies $E_0 \pm \Delta E_a$.
To model ferromagnetic state we assume that 
$\vec{B}_{\rm eff}= (0,0,B_{\rm eff})$, and that electron
states belonging to different bands have opposite spin
orientation along $\hat{z}$ direction,
and opposite orientation of their local orbital moments.
With rising energy of atomic orbitals their radius increases.
Up to the lowest order in energy we get
\begin{equation}
\frac{r_a(E_a)}{\tilde{a}} \, = \,
\frac{r_{0}}{\tilde{a}}
\left( 1 + \kappa \, \frac{E_a - E_0}{w} \right)
\; ,
\end{equation}
where $r_0$ denotes the radius for the orbital state
of the energy $E_0$ and the parameter $\kappa$ represents
how the radius changes with the orbital energy.

\begin{figure}[h]
\includegraphics[angle=0,width=3.3 in]{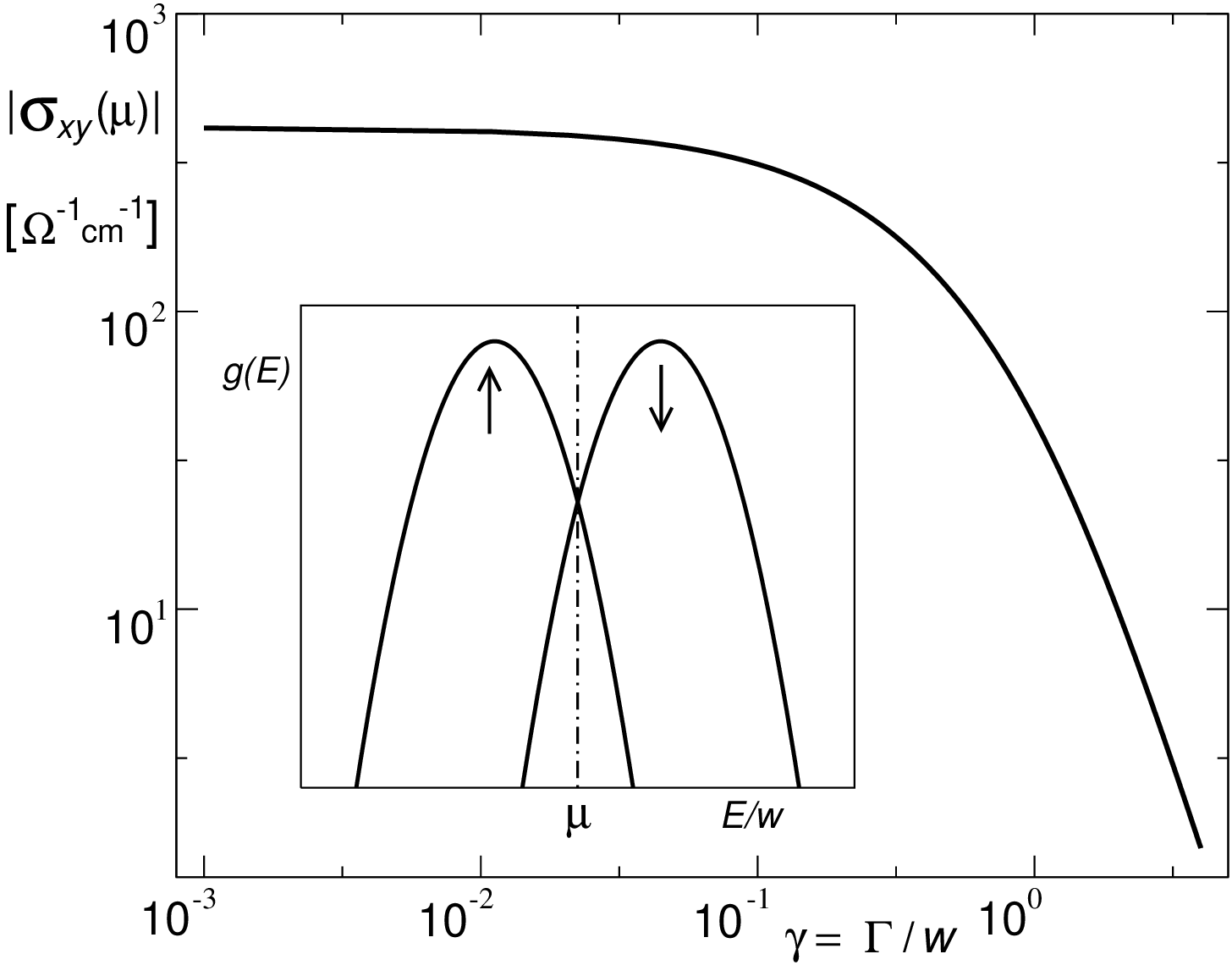}
\caption{Anomalous Hall conductivity for two band model
as a function of the fluctuation strength $\Gamma = \hbar/2\tau$.
Inset shows corresponding density of states.}
\label{sigma_fig}
\end{figure}

The origin of fluctuations modelled by local energy shifts
$\delta$ can vary considerably. In addition to the configurational
disorder like impurities, alloy composition and other types
of the disorder there are thermal fluctuations and that
given by local values of $B_{\rm eff}$.
Assuming that electron hopping between lattice sites is
independent of their band origin, fluctuations of orbital
energies $E_0 \pm \Delta E_a$ can be treated as independent.
Ensemble averaging can thus be performed separately for each
of the band contributions. To get a qualitative estimation
of the effect of fluctuations the Lorentzian distribution of
the parameter $\delta$ will be considered
\begin{equation}
p(\delta,\Gamma) \, = \, \frac{1}{\pi} \,
\frac{\Gamma}{\delta^2 + \Gamma^2}
\; ,
\end{equation}
where the energy parameter $\Gamma$ defines strength of the
fluctuations. For the sake of simplicity so called
virtual crystal approximation \cite{Bedrich} is used
to establish conductivity. In such a case the parameter
$\Gamma$ represents imaginary part of the self-energy
$\Sigma$ entering effective Hamiltonian, Eq.~(\ref{self-energy}),
while the real part of the self-energy equals to zero.
It also defines the electron life time, $\tau=\hbar/2\Gamma$.
Substituting Eq.~(\ref{P_l})
into Eq.~(\ref{sigma_P_n}) for the anomalous Hall conductivity,
and assuming there is one electron per lattice site,
we get
\begin{eqnarray} \label{sigma_tau}
\sigma_{xy}(\mu=E_0) \, = \,
\frac{e^2}{\tilde{a} h} \, \left( \frac{\pi}{2} \right)^{2/3}
\frac{r_0}{\tilde{a}} \times \qquad \qquad \qquad
\nonumber \\ \times
\int\limits_{-1}^{+1}
(1-\xi^2) ( 1 + \kappa \xi)
\left[ p(\xi+\varepsilon, \gamma) -
p(\xi-\varepsilon, \gamma) \right] \, d \xi
\; , \;
\end{eqnarray}
where the half band width $w$ has been used as the energy unit,
$\gamma \equiv \Gamma/w$,
$\varepsilon  \equiv \Delta E_a /w$.
The resulting conductivity dependence on the fluctuation strength 
represented by the parameter $\gamma=\Gamma/w$ is shown on
Fig.~\ref{sigma_fig} for the following set of parameters:
$\tilde{a} = 3$\AA $\,$, $r_{a}(0)/\tilde{a} = 0.3$,
$\kappa = 1$, and $\varepsilon \equiv \Delta E_a/w = 0.6$.

\begin{figure}[h]
\includegraphics[angle=0,width=3.3 in]{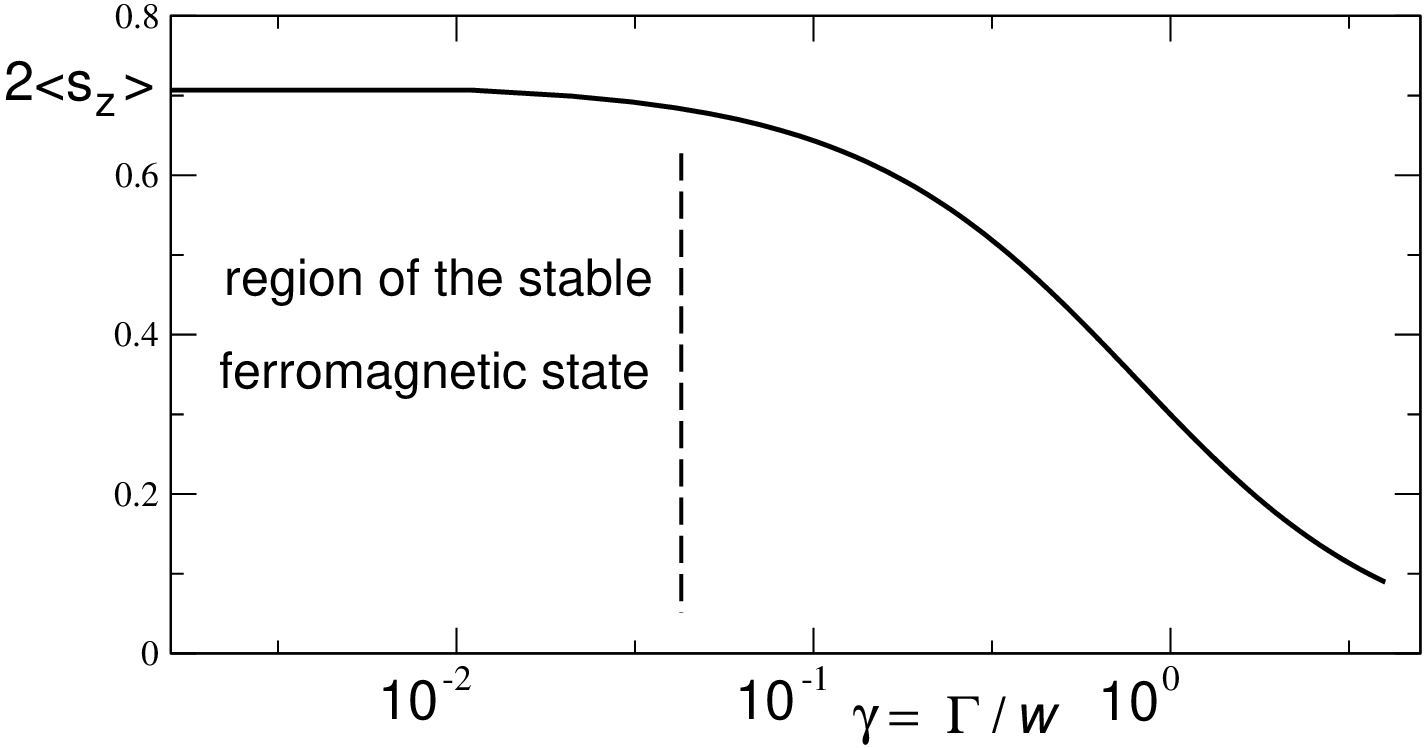}
\caption{Average spin $\langle s_z \rangle$ per site
for two band model as a function of the fluctuation strength
$\Gamma = \hbar/2\tau$. Model parameters are the same as that
used for the dependence presented in Fig.~\ref{sigma_fig}.}
\label{Sz_fig}
\end{figure}

The anomalous Hall conductivity dependence on the fluctuation
strength $\Gamma$ shows the same qualitative features as that
found for multi-$d$-orbital tight-binding model developed by
Kontani et al \cite{Kontani_2}. In the case of a weak disorder,
$\Gamma/\Delta E_a \ll 1$, conductivity is nearly constant
while for strong disorder, $\Gamma/\Delta E_a > 1$,
it decreases with a power of the electron life time
$\tau = \hbar/\Gamma$, which for the present example even
slightly exceeds quadratic dependence. In comparison with
the procedure used by Kontani et al whose evaluation is based
on the conductivity formula including only velocity operators,
presented treatment based on local orbital polarization
moments is much simpler and has a clear interpretation.
Ferromagnetic state, necessary for appearance of anomalous
Hall conductivity, can be characterized by average spin
along magnetization axis which is supposed to be parallel
with $\hat{z}$ direction. In the case of the considered two-band 
model it is assumed that each of them is fully spin-polarized but
in opposite directions, $s_z = \pm 1/2$. If there
are no fluctuations the average spin per
site $\langle s_z \rangle$ reaches a maximum value.
Fluctuations of orbital energies can only lead to suppression
of this value. The dependence of $\langle s_z \rangle$
on the parameter $\Gamma$ representing fluctuation strength
is presented in Fig.~\ref{Sz_fig} for the same set of 
parameters as that used for dependence presented in
Fig.~\ref{sigma_fig}. It shows the same qualitative
features as the dependence of the Hall conductivity on
$\Gamma$. In the region where $\langle s_z \rangle$
is smaller than its maximum value the Hall conductivity
depends on the fluctuation strength. It is the region in
which the disorder is strong enough to weaken ferromagnetic
state.

\begin{figure}[h]
\includegraphics[angle=0,width=3.3 in]{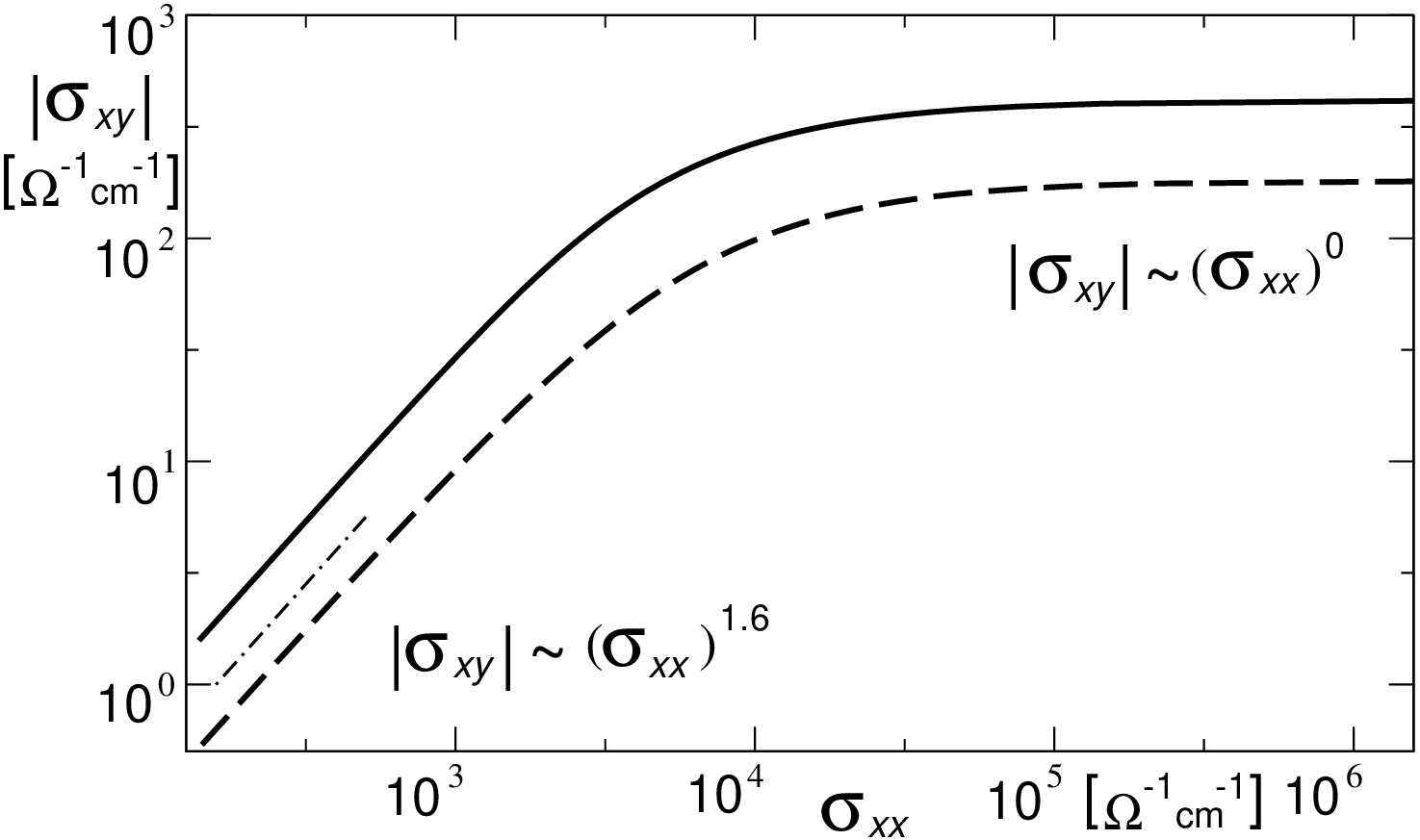}
\caption{Scaling of the Hall conductivity $\sigma_{xy}$
with longitudinal conductivity $\sigma_{xx}$.
Full line is given by the same model parameters
as that used for the dependence presented in
Fig.~\ref{sigma_fig} while for dashed line the parameters
$\kappa$ and $\varepsilon \equiv \Delta E_a/w$ have been changed
($\kappa=0.5$ and $\varepsilon = 0.4$).
Dashed-dotted line shows the slope 1.6.}
\label{sig_scale_fig}
\end{figure}

The simplest approach for diagonal conductivity component
$\sigma_{xx}$, Eq.~(\ref{sigma_ii}), is to neglect vertex
corrections, i.e. to use decoupling
$\langle G v G \rangle_{\rm av} \rightarrow
\langle G \rangle_{\rm av} v \langle G \rangle_{\rm av}$.
In this approximation relaxation time coincides with
electron life time $\tau= \hbar / 2\Gamma$.
Using virtual crystal approximation for the considered
two-band model and probability distribution of fluctuations,
Eq.~(\ref{sigma_ii}), evaluation of $\sigma_{xx}$ becomes trivial.
In Fig.~\ref{sig_scale_fig} obtained scaling of the Hall
conductivity with $\sigma_{xx}$ is shown. It
reveals typical features observed experimentally
\cite{Nagaosa2009,Miyasato}. Especially the case of the 
moderate disorder (good metal regime)
for which $|\sigma_{xy}| \sim (\sigma_{xx})^0$
and that of the strong disorder (bad metal / hopping regime) 
for which $|\sigma_{xy}| \sim (\sigma_{xx})^{1.6}$.

\section{Concluding remarks}
Transport theories of conductivity are traditionally
formulated to give expressions containing velocity matrix
elements only. In the presented treatment the Hall conductivity
has been expressed in terms containing also position operator.
It has been found that only on-energy-shell matrix elements
of both operators are relevant, Eq.~(\ref{sigma_xy_alpha}).
For Bloch electron systems the Hall conductivity is
given by a part of the orbital magnetization of Fermi electrons,
called orbital polarizability momentum,
Eqs.~(\ref{sigma_P_n}) and (\ref{OPM}), which is a quantity
determined by atomic-like orbitals. For a perfect Bloch
system it is equivalent to the expression given by the
Berry phase curvatures, Eq.~(\ref{Bp}).

To test applicability of the derived alternative expression
for the Hall conductivity, Eq.~(\ref{sigma_xy_alpha}),
the two-band model based on the tight-binding approach
has been used. Disorder has been modelled by energy fluctuations
of single site orbitals. It represents intra-band scattering
which can be identified with the so called side-jump scattering.
However, this simple model excludes effect of the skew scattering
since spin of electrons is supposed to be fixed.
Despite of its simplicity it correctly
describes scaling of the anomalous Hall conductivity
$\sigma_{xy}$ with diagonal conductivity component
$\sigma_{xx}$ in the region covering the so called
good metal regime, $|\sigma_{xy}| \sim (\sigma_{xx})^0$,
and the bad metal (hopping) regime,
$|\sigma_{xy}| \sim (\sigma_{xx})^{1.6}$.
The bad metal regime has been identified with the regime
in which disorder becomes strong enough to weaken
ferromagnetic state.
It remains an open question if the presented form of
the general expression for the Hall conductivity in terms
of the local orbitals could also be effective in description
of the skew scattering effect.

To estimate anomalous Hall conductivity for a real
material requires knowledge of local orbitals represented
by Wannier functions and also the specific form of fluctuations
relevant for the studied system. In particular, finite temperature
spin fluctuations are expected to be essential. It is a challenge to
work out such a procedure based on the first principle calculations.
Newly developed numerical techniques allowing
to establish Wannier functions giving the best tight-binding
model parameters \cite{Marzari,Wang} seem to be a proper way
to establish the fluctuation effect upon the anomalous Hall
effect in real materials.

\acknowledgments
The author acknowledges support from
Grant No. GACR 202/08/0551
and the Institutional Research Plan No. AV0Z10100521.
The author thanks Jan Ku\v{c}era for useful comments and
colleagues from Beijing University, Dingping Li for
initializing this work, and Zhongshui Ma for fruitful
discussions.

\appendix\section{}

Periodic part of Bloch functions can be expressed via
Wannier functions $\phi_n(\vec{r}-\vec{R}_l)$ as follows
\begin{equation}
u_{n,\vec{k}}(\vec{r}) \, = \,
\frac{1}{\sqrt{N}} \sum_{l=1}^N
e^{i \vec{k} (\vec{R}_l-\vec{r})} \phi_n(\vec{r}-\vec{R}_l)
\; ,
\end{equation}
where by definition Wannier functions are orthonormal
with respect to their mass-center position vector $\vec{R}_l$.
The expectation value of the position vector $\vec{r}$ reads
\begin{eqnarray}
\vec{r}_n(\vec{k}) \, = \,
\langle n, \vec{k} | \vec{r} \  n, \vec{k} \rangle \, = \,
\qquad \qquad \nonumber \\ =
\frac{1}{2} \int \left[
u^{+}_{n, \vec{k}}(\vec{r}) \, \vec{r} \, u_{n, \vec{k}}(\vec{r}) +
u_{n, \vec{k}}(\vec{r}) \, \vec{r} \, u^+_{n, \vec{k}}(\vec{r})
\right] dr^3  =
\nonumber \\ =
\sum_{l=1}^N
\int u^{+}_{n, \vec{k}}(\vec{r}) \,
\frac{\vec{r} - \vec{R}_l}{2\sqrt{N}} \,
e^{-i \vec{k}(\vec{r} - \vec{R}_l)} \, \phi_n(\vec{r}-\vec{R}_l) \,
d^3r +
\nonumber \\ +
\sum_{l=1}^N
\int u_{n, \vec{k}}(\vec{r}) \,
\frac{\vec{r} - \vec{R}_l}{2\sqrt{N}} \,
e^{i \vec{k}(\vec{r} - \vec{R}_l)} \, \phi^{+}_n(\vec{r}-\vec{R}_l) \,
d^3r +
\nonumber \\ +
\! \sum_{l,l'=1}^N \frac{\vec{R}_l}{2N} \! \int \! \!
e^{i \vec{k}(\vec{R}_l - \vec{R}_{l'})}
\phi_n^+(\vec{r}-\vec{R}_{l'}) \, \phi_n(\vec{r}-\vec{R}_{l}) \,
d^3r + \nonumber
\end{eqnarray}
\begin{eqnarray} +
\! \sum_{l,l'=1}^N \frac{\vec{R}_l}{2N} \! \int \! \!
e^{i \vec{k}(\vec{R}_{l'} - \vec{R}_{l})}
\phi_n(\vec{r}-\vec{R}_{l'}) \, \phi_n^+(\vec{r}-\vec{R}_{l}) \,
d^3r = \nonumber \\ =
\frac{i}{2} \int \! \!
\left( u^{+}_{n, \vec{k}}(\vec{r}) \, \vec{\nabla}_{\vec{k}}
u_{n, \vec{k}}(\vec{r}) -
u_{n, \vec{k}}(\vec{r}) \, \vec{\nabla}_{\vec{k}}
u^{+}_{n, \vec{k}}(\vec{r}) \right) d^3r +
\nonumber \\ +
\frac{1}{N} \sum_{l=1}^N \vec{R}_l
\; , \qquad
\end{eqnarray}
where the last constant term represents the center of mass of the
considered electron system.
For states with $- \vec{k}$ the above relation gives
\begin{equation}
\vec{r}_n( - \vec{k}) =
{\rm Im} \int \! \! u^+_{n, -\vec{k}}(\vec{r}) \,
\vec{\nabla}_{\vec{k}} \, u_{n, -\vec{k}}(\vec{r}) \,
d^3r +
\frac{1}{N} \sum_{l=1}^N \vec{R}_l
\; .
\end{equation}
In the summation over Fermi surface states of the product
$\vec{r}_n(\vec{k}) \, \vec{v}_n(\vec{k})$ entering
Eq.~(\ref{OPM}) the mass-center
position of the electron system is cancelled out
since $\vec{v}_n(\vec{k}) = - \vec{v}_n(-\vec{k})$.
The last term on the right-hand side of the above equation
can thus be excluded from the consideration since
it does not affect the final result.


\begin{thebibliography}{99}
\bibitem{R1_adiabatic_approach}
{Time dependent potential fluctuations are supposed to be so slowly
varying in time that electron states at given time are
determined by the eigenvalue problem defined by the static potential
just equal to that at given time.}
\bibitem{remark_2}
{In the derivation the terms containing the expression
of the form
$\delta(\eta-H)[G^{\pm}(\eta)]^{-1}=
\epsilon (\eta-H\pm i\epsilon)^{-1}$ have been neglected since
they are proportional to infinitesimal $\epsilon$ after
any averaging procedure.}
\end{thebibliography}
\end{document}